\documentclass[11pt]{article}

\usepackage{amsmath}
\usepackage{cite}
\usepackage{amsfonts}
\usepackage{amssymb}
\usepackage{latexsym}
\usepackage{color}
\usepackage[dvipdfmx]{graphicx}
\usepackage{subfigure}
\usepackage{lmodern}
\input{colordvi.tex}

\def\be{\begin{equation}}
\def\ee{\end{equation}}

\begin{document}

\begin{titlepage}
\begin{flushright}
RESCEU-17/16 \\
RUP-16-7\\
YITP-16-43
\end{flushright}
\begin{center}

%\hfill February 2008\\

\vskip .5in

{\Large \bf
Primordial Black Hole Scenario for the Gravitational-Wave Event GW150914 
}
\vskip .45in

{%\large
Misao Sasaki$^{a}$,
Teruaki Suyama$^{b}$,
Takahiro Tanaka$^{c,a}$,
and Shuichiro Yokoyama$^{d}$
}

%\vskip .45in%
{\em
$^a$
   Center for Gravitational Physics, Yukawa Institute for Theoretical Physics,
Kyoto University, Kyoto 606-8502, Japan
}\\

{\em
$^b$
   Research Center for the Early Universe (RESCEU), Graduate School
  of Science,\\ The University of Tokyo, Tokyo 113-0033, Japan
}\\

{\em
$^c$
  Department of Physics, Kyoto University, Kyoto 606-8502, Japan
}\\

{\em
$^d$
  Department of Physics, Rikkyo University, Tokyo 171-8501, Japan
}\\

\end{center}

\vskip .4in

\begin{abstract}
We point out that the gravitational-wave event GW150914 observed by 
the LIGO detectors can 
be explained by the coalescence of primordial black holes (PBHs).
It is found that the expected PBH merger rate would exceed the rate 
estimated by the LIGO Scientific Collaboration and the Virgo Collaboration
if PBHs were the dominant component of dark matter,
while it can be made compatible if PBHs constitute a fraction of dark matter.
Intriguingly, the abundance of PBHs required to explain the suggested lower 
bound on the event rate, $> 2$ events ${\rm Gpc}^{-3} {\rm yr}^{-1}$, roughly coincides with
the existing upper limit set by the nondetection of the cosmic microwave background spectral 
distortion. This implies that the proposed PBH scenario may be tested in the
not-too-distant future.
\end{abstract}
\end{titlepage}

\renewcommand{\thepage}{\arabic{page}}
\setcounter{page}{1}
\renewcommand{\thefootnote}{\#\arabic{footnote}}

%%%%%%%%%%%%%%%
\section{Introduction}
%%%%%%%%%%%%%%%%
The gravitational-wave event GW150914 observed by the LIGO detectors \cite{Abbott:2016blz} revealed 
the existence of black holes (BHs) with a mass of around $30~M_\odot$ in the form of 
binaries.
Although there are several possible explanations for the origin of those BHs as well
as the formation of the binaries (see \cite{TheLIGOScientific:2016htt} and 
references therein), the answer is yet to be elucidated.
Assuming all the BH binaries relevant to the LIGO observation have the same physical 
parameters, such as masses and spins, as those of GW150914, 
the merger event rate was estimated as 
$2-53~{\rm Gpc}^{-3} {\rm yr}^{-1}$ \cite{Abbott:2016nhf}.

In this {\it Letter}, we discuss the possibility that the event GW150914 was 
caused by a merger of a primordial BH (PBH) binary.
PBHs are BHs that have existed since the very early epoch in cosmic 
history before any other astrophysical object had been formed \cite{Zeldovich:1967ei}.
The most popular mechanism to produce PBHs is the direct gravitational collapse of 
a primordial density inhomogeneity \cite{Hawking:1971ei, Carr:1974nx}.
If the primordial Universe were highly inhomogeneous [${\cal O}(1)$
in terms of the comoving curvature perturbation] on superhorizon scales, 
as realized in some inflation models (see \cite{Carr:2009jm} and references therein),
an inhomogeneous region upon horizon reentry would undergo gravitational collapse 
and form a BH. The mass of the BH is approximately equal to the horizon mass at the time of formation,
$M_{\rm BH} \sim 30~M_\odot {\left( \frac{4\times 10^{11}}{1+z_f} \right)}^2$,
where $z_f$ is the formation redshift.
Thus, it is possible that PBHs with a mass of around 
$30~M_\odot$ are formed deep in the radiation-dominated era.

The event rate of the PBH binary mergers has been already given in \cite{Nakamura:1997sm}
for the case where PBHs are massive compact halo objects with their mass around 
$0.5~M_\odot$ and constitute the dominant component of dark matter.
In \cite{Nakamura:1997sm}, it was found that two neighboring PBHs having a sufficiently 
small separation can form a binary in the early Universe and coalesce within the age
of the Universe.
We apply the formation scenario in \cite{Nakamura:1997sm} to the present
 case where the PBHs are 
about $30~M_\odot$ and the fraction of PBHs in dark matter
 is a free parameter.
We present a detailed computation of the event rate in the next section.
The resultant event rate turns out to exceed the event rate mentioned
above ($2-53~{\rm Gpc}^{-3} {\rm yr}^{-1}$) if
PBHs are the dominant component of the dark matter. Intriguingly, however, it
falls in the LIGO range if PBHs are a subdominant component of dark matter
with the fraction that nearly saturates the upper limit set by the
nondetection of the cosmic microwave background (CMB) spectral distortion due to gas 
accretion onto PBHs~\cite{Ricotti:2007au}.

Recently, it was claimed in \cite{Bird:2016dcv} (see also \cite{Clesse:2016vqa}) 
that the event GW150914 as well as the event rate 
estimated by LIGO can be explained by the merger of PBHs even if PBHs
are the dominant component of dark matter.
Our study differs from \cite{Bird:2016dcv} in the following two points: 
(1) the formation process of PBH binaries
and (2) the fraction of PBHs in dark matter.
First, in \cite{Bird:2016dcv} PBH binaries are assumed to be formed
due to energy loss by gravitational radiation when two PBHs accidentally 
pass by each other with a sufficiently small impact parameter.
This mechanism is different from what we consider in this Letter (see the next section).
Second, in \cite{Bird:2016dcv} the fraction of PBHs in dark matter 
to explain the estimated gravitational-wave event rate by the LIGO-Virgo 
Collaboration is of order unity, while
in our case we require it to be as small as the upper limit 
obtained in \cite{Ricotti:2007au}.
Namely, our claim is that PBHs as a small fraction of dark matter
can explain the event rate suggested by the detection of GW150914.

Throughout this Letter, we set the speed of light to be unity, $c=1$.
%%%%%%%%%%%%%%%%%%%%%%%
\section{Event rate of mergers of PBH binaries}
%%%%%%%%%%%%%%%%%%%%%%%

In this section, we estimate the event rate of PBH binary mergers.
We adopt the formation mechanism proposed in \cite{Nakamura:1997sm}
and basically follow the same calculation procedure described in it.
A refined analysis taking into account various effects neglected in
\cite{Nakamura:1997sm} shows that those effects can change the event 
rate estimation at most by $\sim 50\%$ \cite{Ioka:1998nz}.
Given the large uncertainties in the event rate estimated by the LIGO-Virgo Collaboration 
as well as in the upper 
limit on the abundance of PBHs from the nondetection
of the CMB spectral distortion, those corrections are not important and we adopt
the simple method given in \cite{Nakamura:1997sm}.
Our analysis differs from that in \cite{Nakamura:1997sm} in two aspects:
the PBH mass is $30~M_\odot$, and the PBH fraction in dark matter
is a free parameter.
For simplicity, we assume all the PBHs have the same mass.
If necessary, our analysis can be straightforwardly generalized to a realistic 
situation in which the PBH mass function is not monochromatic.

\begin{figure}[tbp]
  \begin{center}
   \includegraphics[width=100mm]{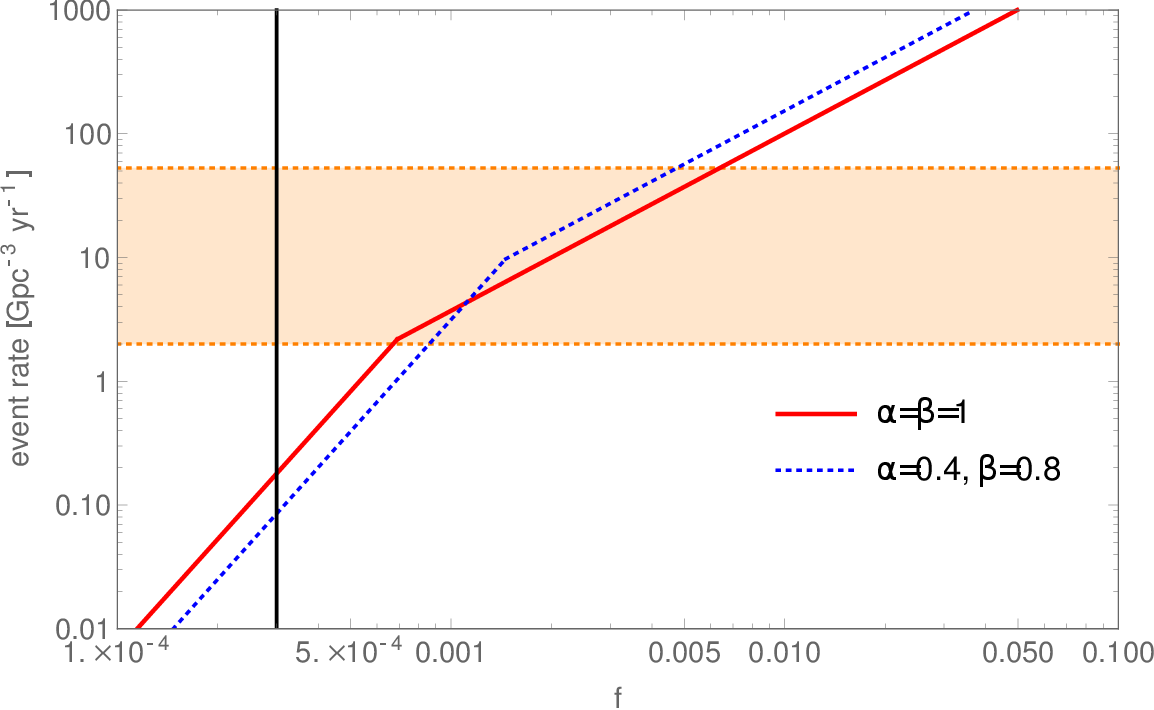}
  \end{center}
  \caption{Event rate of mergers of $30~M_\odot-30~M_\odot$ PBH binaries as
 a function of the PBH fraction in dark matter $f=\Omega_{\rm BH}/\Omega_{\rm DM}$.
 The red line is the case for $\alpha=\beta=1$, which we have employed throughout the calculations.
 The blue dotted line is the case for $\alpha=0.4,~\beta=0.8$ suggested in \cite{Ioka:1998nz}.
 The event rate estimated by the LIGO-Virgo Collaboration is shown as the shaded region colored orange. The black solid line at $f \approx 3\times 10^{-4}$ is the upper limit on $f$ from the
nondetection of the CMB spectral distortion obtained in \cite{Ricotti:2007au}.}
  \label{fig}
\end{figure}

Let $f$ be the fraction of PBHs in dark matter, {\it i.e.},
$\Omega_{\rm BH}=f \Omega_{\rm DM}$.
Then, the physical mean separation ${\bar x}$ of BHs at matter-radiation equality
at the redshift $z=z_{\rm eq}$ is given by
\be
{\bar x}={\left( \frac{M_{\rm BH}}{\rho_{\rm BH}(z_{\rm eq})} \right)}^{1/3}
=\frac{1}{(1+z_{\rm eq}) f^{1/3}} 
{\left( \frac{8\pi G}{3H_0^2} \frac{M_{\rm BH}}{\Omega_{\rm DM}} \right)}^{1/3}.
\ee
Let us consider two neighboring BHs separated by a physical distance $x$ 
at matter-radiation equality.
The pair decouples from the expansion of the Universe and
becomes gravitationally bound when the average energy density of the BHs
over the volume $R^3$, where $R$ is the separation of two BHs, exceeds the
background cosmic energy density $\rho$, that is, when
\be
M_{\rm BH} R^{-3} > \rho (z).
\ee
Using $R=\frac{1+z_{\rm eq}}{1+z}x$, the 
redshift at which the decoupling occurs is given by
\be
\frac{1+z_{\rm dec}}{1+z_{\rm eq}}=f {\left( \frac{\bar x}{x} \right)}^3 -1 > 0.
\ee
This shows that only a pair having $x<f^{1/3} {\bar x}$ can form a binary.
From now on, we require this condition on $x$.
If there are only two BHs on top of the unperturbed Friedmann-Lemaitre-Robertson-Walker Universe, 
after being decoupled from the background expansion,
they move closer together and finally collide without forming a binary.
In a realistic situation, other BHs are also present and the third BH closest 
to the BH pair
affects the infall motion of the BHs in the pair by giving them the tidal force.
As a result, the head-on collision does not happen and the BHs in the pair form a binary
typically having a large eccentricity. 
The major and minor axes of the binary at the formation time are given by 
(denoting by $a$ and $b$, respectively)
\be
a=\frac{\alpha}{f} \frac{x^4}{{\bar x}^3},~~~~~b=\beta {\left( \frac{x}{y} \right)}^3 a, 
\label{ellipse}
\ee
where $y$ is the physical distance to the third BH at $z=z_{\rm eq}$
and $\alpha$ and $\beta$ are numerical factors of ${\cal O}(1)$.
A detailed investigation of the dynamics of the binary formation suggests 
$\alpha = 0.4,~\beta=0.8$ \cite{Ioka:1998nz}. 
In the following analysis, we take $\alpha=\beta=1$ for simplicity.
In Fig.~\ref{fig} showing our estimated merger event rate, 
the event rate in the case $\alpha=0.4,~\beta=0.8$ is also plotted,
which demonstrates that the difference between the two is not significant
compared to the uncertainty of the event rate provided by the LIGO-Virgo Collaboration.

The eccentricity of the binary at the formation time is given by
\be
e=\sqrt{1-{\left( \frac{x}{y} \right)}^6}.
\label{eccentricity}
\ee
By definition, $y>x$ must be satisfied.
In addition to this, we also have the condition $y<{\bar x}$, which yields 
an upper bound on $e$ as
\be
e_{\rm max}=\sqrt{1-f^{3/2} {\left( \frac{a}{\bar x} \right)}^{3/2}}. \label{emax}
\ee
We assume a uniform probability distribution both for 
$x$ and $y$ in three dimensional space \footnote{
If PBHs are formed from the high-$\sigma$ peaks of the random 
Gaussian density fluctuations, the
distribution is not uniform and the PBHs are rather clustered. Intuitively, 
clustering of the PBHs facilitates the
formation of binaries and thus boosts the event rate. In this sense, our 
uniformity assumption would provide the
conservative estimate at least for the case where PBHs originate from
Gaussian density fluctuations.
We thank Jun'ichi Yokoyama for pointing this out to us.}.
Thus, the probability $dP$ that the would-be binary BHs have a separation in $(x,x+dx)$
and that the distance to the perturber BH is in $(y,y+dy)$ is given by
\be
dP=\frac{9}{{\bar x}^6} x^2 y^2 dx dy.
\ee
We can convert this probability distribution function into 
the one for $a$ and $e$ by using the mapping formulae (\ref{ellipse}) 
and (\ref{eccentricity}). The result is given by
\be
dP=\frac{3}{4} f^{3/2} {\bar x}^{-3/2} a^{1/2} e {(1-e^2)}^{-3/2} da de. \label{pdf-ae}
\ee
Once the BHs form a binary, they gradually shrink by gravitational radiation 
and eventually merge. The coalescence time is given 
by \cite{Peters:1963ux, Peters:1964zz}
\be
t=Q a^4 {(1-e^2)}^{7/2},~~~~~Q=\frac{3}{170} {(GM_{\rm BH})}^{-3}.  
\label{coalescing-t}
\ee
Using this equation, we can convert the probability distribution above 
into the one defined in the $t-e$ plane as
\be
dP=\frac{3}{16} {\left( \frac{t}{T} \right)}^{3/8} e {(1-e^2)}^{-\frac{45}{16}} 
\frac{dt}{t} de,~~~~~
T \equiv \frac{{\bar x}^4 Q}{f^4}.
\ee
Integrating this probability density over $e$ for fixed $t$, we obtain the 
probability distribution function for the coalescing time.
The upper limit of $e$ is given by
\be
e_{\rm upper}=\begin{cases}
    \sqrt{1-{\left( \frac{t}{T} \right)}^{\frac{6}{37}}} ~~~~~{\rm for}~t<t_c\\
    \sqrt{1-f^2 {\left( \frac{t}{t_c} \right)}^{\frac{2}{7}}}~~~~~{\rm for}~t\ge t_c,
  \end{cases}
\ee
where $t_c$ is defined by
\be
t_c=Q {\bar x}^4 f^{\frac{25}{3}}.
\ee
The probability that the coalescence occurs in the time interval $(t,t+dt)$ then becomes
\be
dP_t=\begin{cases}
    \frac{3}{58} \bigg[ -{\left( \frac{t}{T} \right)}^{3/8}
+{\left( \frac{t}{T} \right)}^{3/37} \bigg] \frac{dt}{t}~~~~~{\rm for}~t<t_c\\
    \frac{3}{58} {\left( \frac{t}{T} \right)}^{\frac{3}{8}} \bigg[
    -1+{\left( \frac{t}{t_c} \right)}^{-\frac{29}{56}} f^{-\frac{29}{8}} \bigg] 
    \frac{dt}{t}~~~~~{\rm for}~t\ge t_c.
  \end{cases}\label{dpt}
\ee
The probability that the coalescence happens within the time interval $(0, t)$ is then
simply given by $P_c(t)=\int_0^t dP_t$.
The LIGO-Virgo Collaboration obtained the event rate $2-53~{\rm Gpc}^{-3} {\rm yr}^{-1}$ for the BH binary 
coalescence from the observation of the event GW150914
at $z=z_{\rm GW150914}=0.09$ \cite{Abbott:2016nhf}.
It is not a trivial task to compare the event rate of PBH coalescence 
with that given by the LIGO-Virgo Collaboration in a rigorous manner
since the event rate is assumed to be uniform in comoving 
volume and source time in \cite{Abbott:2016nhf}
while this is not true in our case.
Here we simply ignore the effects of cosmological evolution and consider the
event rate evaluated at the present time, which is obtained by taking the limit
 $\lim_{\Delta t \to 0} \frac{P_c(t_0)-P_c(t_0-{\Delta t})}{\Delta t}$,
where $t_0$ is the age of the Universe.
Thanks to the relatively low value of $z_{\rm GW150914}$, this approximation 
is valid within the accuracy we are care about.
Indeed, changing $\Delta t=0$ to $\Delta t$ 
corresponding to the average redshift of the observed volume, ${\bar z} \approx 0.15$,
shifts the event rate only by less than $25 \%$.

The present average number density of PBHs $n_{\rm BH}$ is given by
\be
n_{\rm BH}=\frac{3H_0^2}{8\pi G} \frac{\Omega_{\rm BH}}{M_{\rm BH}}.
\ee
Then, the event rate becomes
\be
{\rm event~rate}=n_{\rm BH}\lim_{\Delta t \to 0} \frac{P_c(t_0)-P_c(t_0-{\Delta t})}{\Delta t}=
\frac{3H_0^2}{8\pi G} \frac{\Omega_{\rm BH}}{M_{\rm BH}} 
\frac{dP_c}{dt}\bigg|_{t_0}. \label{event-rate}
\ee
Figure~\ref{fig} shows the event rate (\ref{event-rate}) as a function of $f$.
We adopt $\Omega_{\rm DM}=0.27$ and $H_0=70~{\rm km/Mpc/s}$.
We find that the event rate falls in the LIGO range if
the fraction $f$ is around $10^{-3}$.
As a comparison, we also show the case with $\alpha=0.4,~\beta=0.8$ [see Eq.~(\ref{ellipse})]. 
Given that our event rate is derived from the order of magnitude argument,
the difference between the two cases is reasonably acceptable.
Furthermore, the difference of the event rates for any $f$ is smaller than the uncertainty 
of the event rate by the LIGO-Virgo Collaboration, which also justifies our simplified analysis.

%%%%%%%%%%%%%%%%%% 
\section{Discussion}
%%%%%%%%%%%%%%%%%%

It is quite intriguing that the event rate of mergers of PBH binaries falls into 
the range of that given by the LIGO-Virgo Collaboration
when the fraction of PBHs in dark matter nearly saturates
the upper limit obtained by the nondetection of the CMB spectral 
distortion~\cite{Ricotti:2007au}. In other words, our PBH scenario
may be experimentally falsifiable in the not-too-distant future.
It should be noted that the upper limit hinges on the various approximations
made to deal with the complexity of the accretion process and it is not 
an easy task to quantify
how uncertain the result of \cite{Ricotti:2007au} is.
Because of this, we consider the upper limit as the order of magnitude estimation.
Furthermore, it has been assumed that PBHs initially distribute uniformly in space.
Depending on the statistical properties of the primordial perturbations,
this assumption is not necessarily satisfied and it is possible that PBHs upon formation are clustered.
In the latter case, it is expected that the binary formation becomes more efficient
and the merger event rate is enhanced compared to the present case. 
Another potentially important effect that we did not take into account is the dynamical friction 
acting on the BHs in the binaries caused by the interaction with dark matter
trapped in the gravitational potential of the binaries.
If the PBH fraction $f$ is as small as the value corresponding to the kink
in Fig.~\ref{fig}, the mass of the trapped dark matter 
becomes comparable to the BH mass at the matter-radiation equality and grows further
in the matter dominated era. 
Since the time scale of the dynamical friction is much shorter than the age of the Universe,
it may be possible that the binary size quickly changes by a factor of ${\cal O}(1)$.
Quantifying this effect on the event rate is beyond the scope of this paper
(see \cite{Hayasaki:2009ug} for the related discussion).
With the coincidence between our estimated event rate and the observation 
within the uncertainties mentioned above, 
we conclude that the event GW150914 could be a PBH binary merger.

Let us briefly mention that it is unlikely that the PBH binary is disrupted by other compact objects such as other PBHs and stars. 
The typical major axis of the PBH binary for a given life time of the binary,
which we take to be the age of the Universe $t_0$, is given
as a solution of $t_0=Q a^4 {(1-e_{\rm max}^2)}^{7/2}$ since
the possible largest eccentricity is the most probabilistically favored.
We then find that 
$a \approx 7\times 10^4~{\rm AU} {\left( \frac{f}{f_c} \right)}^{-28/37}$ for $f \ge f_c$
and $a \approx 7\times 10^4~{\rm AU}$ for $f \le f_c$,
where $f_c \approx 7\times 10^{-4}$ is $f$ at the kink in Fig.~\ref{fig}.
Since the probability that a given PBH binary is disrupted by the
compact objects becomes smaller for smaller $f$ if $f<f_c$, we now focus on
$f \ge f_c$. 
The PBH binary will be disrupted if the velocity gain of the PBH due to the gravitational
force by the incident compact object becomes comparable to the orbital velocity of
the binary. 
Denoting by $d$ the closest distance that the compact object approaches the PBH
in the binary, 
the velocity gain is roughly estimated as $G m/(vd)$,
where $v$ is the typical relative velocity between the binary and the compact object
and $m$ is the mass of the compact object.
Then, the maximum $d$ for which disruption occurs is written as $d_{\rm max} \approx a(m/M_{\rm BH}) (v_{\rm orbital}/v)$,
where $v_{\rm orbital}$ is the velocity of the PBHs in the binary.
Thus, the probability that a given PBH binary collides with other compact objects 
within the age of the Universe is estimated as 
$P \sim d_{\rm max}^2 n v t_0$,
where $n$ is the number of the compact objects considered.
Using this formula, for the PBH binary residing in dark matter halos like the Milky Way,
we have 
\be
P \sim  7\times 10^{-8} {\left( \frac{v}{200~{\rm km/s}} \right)}^{-1}
\left( \frac{M_{\rm halo}}{10^{12}M_\odot} \right) 
{\left(\frac{L_{\rm halo}}{100~{\rm kpc}} \right)}^{-3}
{\left( \frac{f}{f_c} \right)}^{\frac{9}{37}} {\left( \frac{m}{30~M_\odot} \right)}^2,
\ee
where $M_{\rm halo}$ and $L_{\rm halo}$ are the halo mass and halo size,
respectively.
Thus, such PBH binaries are not likely to be disrupted.
If the PBH binary resides in a stellar environment like a galactic disk,
the disruption probability is
\be
P \sim 3\times 10^{-3} {\left( \frac{v}{200~{\rm km/s}} \right)}^{-1}
{\left( \frac{f}{f_c} \right)}^{-\frac{28}{37}}
\left( \frac{n_{\rm star}}{1~{\rm pc}^{-3}} \right)
{\left( \frac{m}{1~M_\odot} \right)}^2.
\ee
Thus, such PBH binaries are also likely to survive for the age of the Universe.
From these estimations, we conclude that most PBH binaries are not disrupted by
encounters with other compact objects.

At present, we do not know how to discriminate the PBH scenario from other 
astrophysical scenarios (see, {\it e.g.}, \cite{TheLIGOScientific:2016htt, O'Leary:2016qkt}).
For instance, a scenario based on Population III binaries also explains 
the high event rate with the peak of the BH mass distribution 
around $\sim 30~M_\odot$ \cite{Kinugawa:2014zha, Kinugawa:2015nla} (see also \cite{Hartwig:2016nde}).
However, as we mentioned in the above, the approximate coincidence of
the estimated fraction of PBHs in dark matter with the upper limit from
the absence of CMB spectral distortions implies that our PBH scenario
may be experimentally proved in the near future. 
In particular, the proposed experiment PIXIE \cite{Kogut:2011xw} is 
supposed to measure the CMB spectral distortion down to the level of $10^{-8}$,
which is roughly a 4 orders of magnitude improvement from the current sensitivity.
Since the CMB spectral distortion produced by the PBHs is proportional 
to the fraction $f$, the upper limit on $f$ will be also improved 
by 4 orders of magnitude if no spectral distortion is detected.
Thus, if the PBH scenario proposed in this Letter turns out to be the case,
future CMB experiments should detect 
the spectral distortion with great significance,
which is a distinctive feature of this scenario.
On the other hand, if that were not to happen, 
the PBH scenario would be strongly disfavored.

Another potentially useful method to discriminate the PBH scenario from the others
is to exploit the distribution of two BH masses in binaries.
A number of detections of BH mergers, which will occur true in the coming years,
will bring us information about the distribution of the binary parameters.
It was demonstrated in \cite{O'Leary:2016qkt} that the binary distribution may be used to
reconstruct the initial mass function of BHs in an astrophysical scenario where 
the mergers occur in dense stellar systems.
It is worth studying the binary distribution in the PBH scenario and it will be 
interesting to see if a distinct feature specific to this scenario appears.

Last, but not least, another interesting direction to pursue is to investigate the low frequency
stochastic gravitational-wave background
continuously emitted by binary PBHs that are still in the phase of 
orbital motion \cite{Ioka:1998gf}.
We naturally expect the presence of an enormous number of such PBH binaries 
in the present Universe. 
It would be interesting to determine the spectrum of such gravitational waves and 
clarify whether it helps to test our PBH scenario.

%%%%%%%%%%%%%%%%%%%%%%%%%%
\section*{Acknowledgments}
%%%%%%%%%%%%%%%%%%%%%%%%%%
We thank Dr.~Ryan Magee for helpful comment.
We thank the organizers and participants of the mini-workshop on inflation
 at the Yukawa Institute 
for Theoretical Physics, YITP-X-15-5 (March 2016), where this work was initiated.  
We would like to thank Bernard Carr, Takeshi Chiba, Yousuke Itoh, Bence Kocsis, Yuuiti Sendouda and 
Jun'ichi Yokoyama for useful discussions.
This work was supported by MEXT KAKENHI Nos.~15H05888, 15K21733,
JSPS Grant-in-Aid for Young Scientists (B) No.15K17632
(T.S.) and No.15K17659 (S.Y.), 
the Grant-in-Aid for Scientific Research No. 26287044 (T.T.),
MEXT Grant-in-Aid for Scientific Research on
Innovative Areas, "New Developments in Astrophysics
Through Multi-Messenger Observations of Gravitational
Wave Sources", Nos. 24103001, 24103006 (T.T.), No.15H00777 (T.S.), 
and by the Grant-in-Aid from the Ministry of Education, Culture, Sports, 
Science and Technology (MEXT) of Japan No. 15H02087 (T.T.).

\bibliographystyle{unsrt}
\bibliography{draft}

\end{document}